# Population Dynamics and the Optical Absorption in Hybrid Metal Nanoparticle ─ Semiconductor Quantum dot Nanosystem

Nam-Chol Kim,[†,*] Chung-Il Choe,[†] Myong-Chol Ko,[†] Gwang Hyok So,[†] Il-Gwang Kim[††]

[†] Department of Physics, Kim Il Sung University, Pyongyang, DPR. Korea

[††]Natural Science Academy, Kim Il Sung University, Pyongyang, DPR. Korea

**Abstract:** We studied theoretically the population dynamics and the absorption spectrum of hybrid nanosystem consisted of a matal nanoparticle (MNP) and a semiconductor quantum dot(SQD). We investigated the exciton-plasmon coupling effects on the population dynamics and the absorption properties of the MNP-SQD hybrid nanostructure. Our results show that the nonlinear optical response of the hybrid nanosystem can be greatly enhanced or depressed due to the exciton-plasmon couplings.

[*] Electronic mail: ryongnam19@yahoo.com

# 1. Introduction

   With the rapid development of nanotechnology, the hybrid nanostructures based on metal nanoparticles (MNPs) and semiconductor quantum dots (SQDs) are the area of considerable current interest[1,2]. The MNP-SQD hybrid systems exhibit many interesting phenomena such as plasmon-induced fluorescence enhancement quenching [3], plasmon-assisted Forster energy transfer [4], generation of a single plasmons, induced exciton-plasmon-photon conversion, modifying the spontaneous emission in semiconductor quantum dots (SQDs) [5], etc. Because of their extraordinary properties such as enhancement of radiative emission rates, absorption of light and nonradiative energy transfer, the hybrid nanosystem coupled with quantum dots have been intensively studied. The MNP-SQD hybrid system exhibits the third-order optical nonlinearity [6,7]. It was also shown that dark plasmon-exciton hybridization strengthens the nonlinear light-quantum emitter interaction and provides a novel means to control optical properties at the nanoscale [8,9,10]. The hybrid nanostructure is obviously different from conventional low-dimensional semiconductors. It is very important to investigate the optical nonlinearity in a hybrid nanosystem. More significant attentions have been focused on the emerging field of quantum plasmonics with the goal of making devices for quantum information processing [11] as single-photon transistors [12] or quantum switch [13]. Additionally, the bistability of the hybrid nanosystem [14], the dynamics of exciton populations in a $\Lambda$-type [5] and ladder-type [15] SQD close to a metallic nanorod has been considered previously. However, in the previous research work, most efforts focus on the hybrid nanosystem consisted of a metal nanoparticle and the two-level energy structure of SQD.

   In this Letter, we will study theoretically the exciton-plasmon coupling effects on the population dynamics and absorption in the hybrid MNP-SQD system, where SQD has V-type three-level structure. The basic excitations in the MNP are the surface plasmons with a continuous spectrum. In SQDs, the excitations are the discrete interband excitons. When the exciton energy in a SQD lies in the vicinity of the plasmon peak of the MNP, the coupling of the plasmon and exciton becomes very strong. The exciton coherent dynamics are modified strongly by the strong coherent interaction.

## 2. Model and Theory

We will study a hybrid molecule composed of a spherical MNP of radius $a$ and a spherical SQD with radius $r$, where the center-to-center distance between the two nanoparticles is $R$ (Fig. 1(a)). We consider a SQD with three-level V-type structure, the three excitonic states of which are denoted as $|1>$, $|2>$, and $|3>$, with energies $\hbar\omega_j$ ($j=1, 2, 3$) as shown in Fig. 1(b). The transition frequencies between the upper levels $|2>$ and $|3>$ and the ground level $|1>$ are $\omega_{21}$ and $\omega_{31}$, respectively. We also denote that $\omega_{31} - \omega_{21} = \omega_{32} \equiv 2\Delta$, where $\omega_{32}$ is the frequency separation of the excited levels. The upper levels $|2>$ and $|3>$ decay to the ground level $|1>$ are denoted as $\gamma_{21}$ and $\gamma_{31}$, respectively.

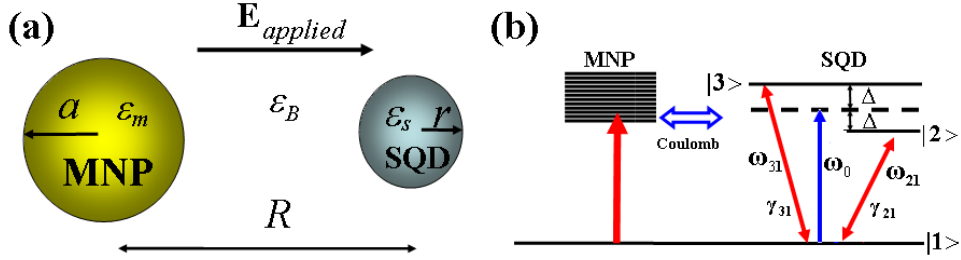

Fig. 1 (Color online). (a) Schematic diagram of the MNP-SQD hybrid system. $r$ and $a$ are the radii of SQD and MNP, respectively. $R$ is the center-to-center distance between SQD and MNP. $\varepsilon_B$, $\varepsilon_s$ and $\varepsilon_m$ are the dielectric constants of the background medium, SQD and MNP, respectively. (b) Energy level diagram of the MNP-SQD hybrid system. Here, $\omega_{31}-\omega_{21}=\omega_{32}\equiv 2\Delta$. The upper levels $|2>$ and $|3>$ decay to the ground level $|1>$ are denoted as $\gamma_{21}$ and $\gamma_{31}$, respectively.

A laser field with frequencies $\omega_{12}$ and $\omega_{13}$ couples the ground state $|1>$ and the two excited states $|2>$ and $|3>$, which is given as follows

$$\vec{E} = E_{21}\vec{u}_{21}\cos(\omega_{21}t) + E_{31}\vec{u}_{31}\cos(\omega_{31}t) = \frac{1}{2}E_{21}\vec{u}_{21}(e^{i\omega_{21}t}+e^{-i\omega_{21}t}) + \frac{1}{2}E_{31}\vec{u}_{31}(e^{i\omega_{31}t}+e^{-i\omega_{31}t})$$

where $E_{21}$ ($E_{31}$) is the slowly varying amplitude along the direction $\vec{u}_{21}$ ($\vec{u}_{31}$) of the field driving the transition $|1\rangle \leftrightarrow |2\rangle$ ($|1\rangle \leftrightarrow |3\rangle$). For simplicity, we set $E_{21}=E_{31}=E_0$.

The Hamiltonian of the hybrid system in the rotating-wave approximation can be written as

$$H = \hbar\sum_{j=1}^{3}\omega_j \hat{a}_j^+ \hat{a}_j - \mu_{21}E_{SQD}^{21}(\sigma_{21}+\sigma_{12}) - \mu_{31}E_{SQD}^{31}(\sigma_{31}+\sigma_{13}), \quad (1)$$

where $\hat{a}_j^+$ and $\hat{a}_j$ are the creation and annihilation operators of the $j$th exciton, respectively, and $\sigma_{ij} = |i\rangle\langle j|$ is the dipole transition operator between $|i>$ and $|j>$ of SQD. $\mu_{j1}(j=2,3)$ is the interband dipole matrix element of the excitonic transitions between $|j>$ and $|1>$. $E_{SQD}^{21}$ and $E_{SQD}^{31}$ are the field felt by SQD driving the transitions $|1\rangle \leftrightarrow |2\rangle$ and $|1\rangle \leftrightarrow |3\rangle$, respectively,

$$E_{SQD}^{j1} = \frac{1}{\varepsilon_{eff}}\left(E + \frac{1}{4\pi\varepsilon_B}\frac{S_\alpha^{j1} P_{MNP}^{j1}}{R^3}\right), \qquad (2)$$

where $S_\alpha^{j1}$ ($j=2,3$) is equal to 2 (−1) when $\vec{u}_{j1}$ is parallel(perpendicular) to the $z$ ($x$, $y$) axis. The z direction corresponds to the axis of the hybrid system and $\varepsilon_{effS} = (2\varepsilon_B + \varepsilon_s)/3\varepsilon_B$ accounts for the screening of the dielectric material. The dipole $P_{MNP}^{j1}$ ($j=2,3$) comes from the charge induced on the surface of the MNP, and depends on the total field due to the SQD as $P_{MNP}^{j1} = (4\pi\varepsilon_B)a^3[\gamma(\omega)\tilde{E}_{MNP}^{j1(+)}e^{-i\omega t} + \gamma(\omega)^*\tilde{E}_{MNP}^{j1(-)}e^{i\omega t}]$, where $E_{MNP}^{j1(\pm)}$ ($j=2,3$) are the positive and negative frequency parts of the electric field felt by MNP. The total field acting on the MNP by the driving field $E_{j1}$, $E_{MNP}$, is just

$$E_{MNP}^{j1} = \left(E + \frac{1}{4\pi\varepsilon_B}\frac{s_\alpha^{j1} P_{SQD}^{j1}}{\varepsilon_{effs} R^3}\right), \quad (j=2,3) \qquad (3)$$

and $\gamma(\omega) = (\varepsilon_m(\omega) - \varepsilon_B)/[2\varepsilon_B + \varepsilon_m(\omega)]$ is the dipole polarizability of the MNP. The dipole $P_{SQD}^{j1}$ ($j=2,3$) can be written via the off-diagonal elements of the density matrix as $P_{SQD}^{j1} = \mu_{j1}(\rho_{j1} + \rho_{1j})$ ($j=2,3$). Factoring out the high frequency time dependence of the off-diagonal terms of the density matrix, we can write $\rho_{j1} = \tilde{\rho}_{j1}e^{i\omega_{j1}t}$, $\rho_{1j} = \tilde{\rho}_{1j}e^{-i\omega_{1j}t}$ ($j=2,3$). Putting the above relations into $E_{MNP}^{j1}$, we obtain

$$E_{MNP}^{j1} = \left(\frac{E}{2} + \frac{1}{4\pi\varepsilon_B}\frac{s_\alpha^{j1}\mu_{j1}}{\varepsilon_{effs}R^3}\tilde{\rho}_{1j}\right)e^{-i\omega_{j1}t} + \left(\frac{E}{2} + \frac{1}{4\pi\varepsilon_B}\frac{s_\alpha^{j1}\mu_{j1}}{\varepsilon_{effs}R^3}\tilde{\rho}_{j1}\right)e^{i\omega_{j1}t} \qquad (4)$$

With using the above relation (5), we can rewrite $P_{MNP}^{j1}$ ($j=2,3$) as

$$P_{MNP}^{j1} = (4\pi\varepsilon_B)a^3\left[\gamma(\omega)\left(\frac{E}{2} + \frac{1}{4\pi\varepsilon_B}\frac{s_\alpha^{j1}\mu_{j1}}{\varepsilon_{effs}R^3}\tilde{\rho}_{1j}\right)e^{-i\omega_{j1}t} + \gamma(\omega)^*\left(\frac{E}{2} + \frac{1}{4\pi\varepsilon_B}\frac{s_\alpha^{j1}\mu_{j1}}{\varepsilon_{effs}R^3}\tilde{\rho}_{j1}\right)e^{i\omega_{j1}t}\right]$$

(5)

Substituting the equation (5) into the relation (2), we can write the following equations as

$$E_{SQD}^{j1} = \frac{1}{\varepsilon_{effs}}\left(E + \frac{1}{4\pi\varepsilon_B}\frac{S_\alpha^{j1}P_{MNP}^{j1}}{R^3}\right) = \frac{1}{\varepsilon_{effs}}\left\{\frac{1}{2}Ee^{i\omega_{j1}t} + \frac{1}{2}Ee^{-i\omega_{j1}t}\right.$$
$$\left. + \frac{S_\alpha^{j1}a^3}{R^3}\left[\gamma(\omega)\left(\frac{E}{2} + \frac{1}{4\pi\varepsilon_B}\frac{s_\alpha^{j1}\mu_{j1}}{\varepsilon_{effs}R^3}\tilde{\rho}_{1j}\right)e^{-i\omega_{j1}t} + \gamma(\omega)^*\left(\frac{E}{2} + \frac{1}{4\pi\varepsilon_B}\frac{S_\alpha^{j1}\mu_{j1}}{\varepsilon_{effs}R^3}\tilde{\rho}_{j1}\right)e^{i\omega_{j1}t}\right]\right\}, \quad (6)$$
$$= \frac{\hbar}{\mu_{j1}}\left\{(\Omega_{j1} + G_{j1}\tilde{\rho}_{j1})e^{-i\omega_{j1}t} + (\Omega_{j1}^* + G_{j1}^*\tilde{\rho}_{j1})e^{i\omega_{j1}t}\right\}$$

where $G_{j1} = \dfrac{(s_\alpha^{j1})^2 \gamma(\omega)a^3\mu_{j1}^2}{4\pi\varepsilon_B\hbar\varepsilon_{effs}^2 R^6}$ and $\Omega_{j1} = \dfrac{E\mu_{j1}}{2\hbar\varepsilon_{effs}}\left(1 + \dfrac{\gamma(\omega)a^3 s_\alpha^{j1}}{R^3}\right)$ $(j=2,3)$. $\Omega_{j1}$ is the

normalized Rabi frequency associated with the external field and the field produced by the induced dipole moment $P_{MNP}^{j1}$ of the MNP [16]. The first term in $\Omega_{j1}$ is just the Rabi frequency resulting from the direct coupling to the applied field and the second term is the Rabi frequency resulting from the field produced by the dipole moment of the MNP induced by the applied field. On the other hand, $G_{j1}$ comes from the interaction between the polarized SQD and the MNP. More specifically, $G_{j1}$ comes from when the applied field polarizes the SQD, which in turn polarizes the MNP and then produces a field to interact with the SQD. As we can see easily, $G_{j1}$ is proportional to $\mu^2$, thus it can be analyzed as the self-interaction of the SQD because this coupling to the SQD depends on the polarization of the SQD.

Now, we can solve the master equation,

$$\frac{d\hat{\rho}}{dt} = -\frac{i}{\hbar}[\hat{H},\hat{\rho}] + L(\hat{\rho}), \qquad (6)$$

where $\Gamma(\rho)$ is the relaxation matrix defined as

$$L(\hat{\rho}) = \frac{1}{2}[\gamma_{21}(2a_1\rho a_1^+ - a_1^+a_1\rho - \rho a_1^+a_1) + \gamma_{31}(2a_2\rho a_2^+ - a_2^+a_2\rho - \rho a_2^+a_2) \quad (7)$$

Making use of the rotating wave approximation, we arrive at a set of coupled equations.

$$\frac{d\rho_{11}}{dt} = -i\Omega_{21}\tilde{\rho}_{21} - iG_{21}\tilde{\rho}_{12}\tilde{\rho}_{21} + i\Omega_{21}^*\tilde{\rho}_{12} + iG_{21}^*\tilde{\rho}_{21}\tilde{\rho}_{12}$$
$$- i\Omega_{31}\tilde{\rho}_{31} - iG_{31}\tilde{\rho}_{13}\tilde{\rho}_{31} + i\Omega_{31}^*\tilde{\rho}_{13} + iG_{31}^*\tilde{\rho}_{31}\tilde{\rho}_{13} - \gamma_{21}\rho_{22} - \gamma_{31}\rho_{33}$$

$$\frac{d\rho_{22}}{dt} = i\Omega_{21}\tilde{\rho}_{21} + iG_{21}\tilde{\rho}_{12}\tilde{\rho}_{21} - i\Omega_{21}^*\tilde{\rho}_{12} - iG_{21}^*\tilde{\rho}_{21}\tilde{\rho}_{12} - \gamma_{21}\rho_{22}$$

$$\frac{d\rho_{33}}{dt} = i\Omega_{31}\tilde{\rho}_{31} + iG_{31}\tilde{\rho}_{13}\tilde{\rho}_{31} - i\Omega_{31}^*\tilde{\rho}_{13} - iG_{31}^*\tilde{\rho}_{31}\tilde{\rho}_{13} - \gamma_{31}\rho_{33} \quad (8)$$

$$\frac{d\tilde{\rho}_{12}}{dt} = i(\omega - \omega_2 + \omega_1)\tilde{\rho}_{12} + i(\Omega_{21} + G_{21}\tilde{\rho}_{12})(\rho_{11} - \rho_{22}) - i(\Omega_{31} + G_{31}\tilde{\rho}_{13})\tilde{\rho}_{23} - \frac{1}{2}\gamma_{21}\tilde{\rho}_{12}$$
$$\frac{d\tilde{\rho}_{13}}{dt} = i(\omega - \omega_3 + \omega_1)\tilde{\rho}_{13} + i(\Omega_{31} + G_{31}\tilde{\rho}_{13})(\rho_{11} - \rho_{33}) - i(\Omega_{21} + G_{21}\tilde{\rho}_{12})\tilde{\rho}_{23} - \frac{1}{2}\gamma_{31}\tilde{\rho}_{13}$$
$$\frac{d\tilde{\rho}_{23}}{dt} = i(\omega_2 - \omega_3)\tilde{\rho}_{23} - i(\Omega_{21}^* + G_{21}^*\tilde{\rho}_{21})\rho_{13} + i(\Omega_{31} + G_{31}\tilde{\rho}_{13})\rho_{21} - \frac{1}{2}(\gamma_{31} + \gamma_{21})\tilde{\rho}_{23}$$

where $\rho_{23} = \tilde{\rho}_{23}$, $\rho_{32} = \tilde{\rho}_{32}$. $\gamma_{31}$ and $\gamma_{21}$ represent the radiative decay rates of the excitation states $|3\rangle$ and $|2\rangle$ due to spontaneous emission, respectively. For simplicity, we assume that $\mu_{21} = \mu_{31} = \mu$.

## 3. Numerical Results and Discussions

We now consider the exciton-plasmon coupling effects on the population and the optical absorption of hybrid MNP-SQD system numerically, where the SQD has V-type three-level structure. The SQD considered in our model can be a single self-assembled $In_{0.5}Ga_{0.5}As/GaAs$ SQD[17], where the fine-structure states |2> and |3> of an exciton define a V-type three-level system composed of two orthogonal transition dipole moments. These excitonic states originate in the shape anisotropy of the SQD and reveal an interesting property in Rabi oscillations and Raman beats. The numerical values of the parameters used in our calculations are set as follows; $\varepsilon_s = 12.96$, $\varepsilon_B = 6$, $\hbar\omega_0 = 1.38\text{eV}$, $\hbar\Delta = 70\mu\text{eV}$, $\gamma_{21} \approx \gamma_{31} = 1.25\,\text{ns}^{-1}$.

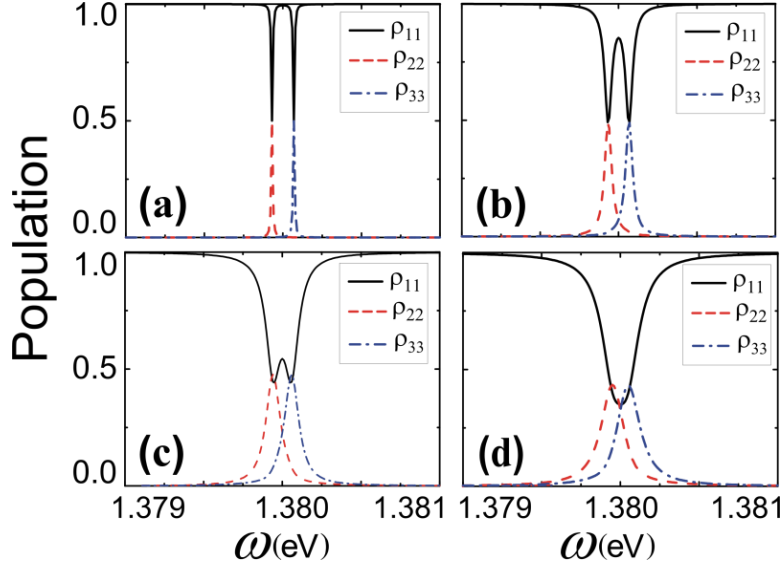

Fig. 2 (Color online). Populations of the SQD levels versus the frequency of the exciting laser field, $\omega$. (a) I=0.1W/cm$^2$, r=0.65nm. (b) I=1W/cm$^2$, r=0.65nm. (c) I=10W/cm$^2$, r=0.65nm. (d) I=1W/cm$^2$, r=3nm. Here, we set R=13nm, a=3nm, $S_\alpha^{12}=2$ and $S_\alpha^{13}=2$.

We first consider the influence of the frequency of the exciting laser field on the population of the V-type three-level system. Figure 2 shows the population dynamics of the V-type three-level QD in the MNP-SQD hybrid system. Figures 2(a), 2(b) and 2(c) display the variations of the distrdistribution of the populations of the different levels for the SQD of the hybrid nanosystem. As we can see easily, there appears a peak of the population $\rho_{11}$ between the two maximum peaks of the excited states |2> and |3>. The height of the peak of the population of the ground state |1> becomes lower as the intensity of the applied field becomes strong. On the other hand, the widths of the peaks of the populations of the excited states |2> and |3> become wider as the intensity of the applied field becomes strong as shown in Figs. 2(a)~2(c). We can also consider the influence of the radius of SQD on the population dynamics. Figures 2(b) and 2(d) show the influence of the radius of SQD on the population dynamics. The width and the height of the population peaks becomes wider and lower, respectively, as the radius of SQD becomes larger. We can also find that there can be disappeared the peak of the population of the ground level |1> as shown in Fig. 2(d). From the results shown in Fig. 2, we found that one can control the population dynamics of the SQD, for example, including the width and height of the populations by adjusting the parameters such as the intensity of applied field and the radius of SQD.

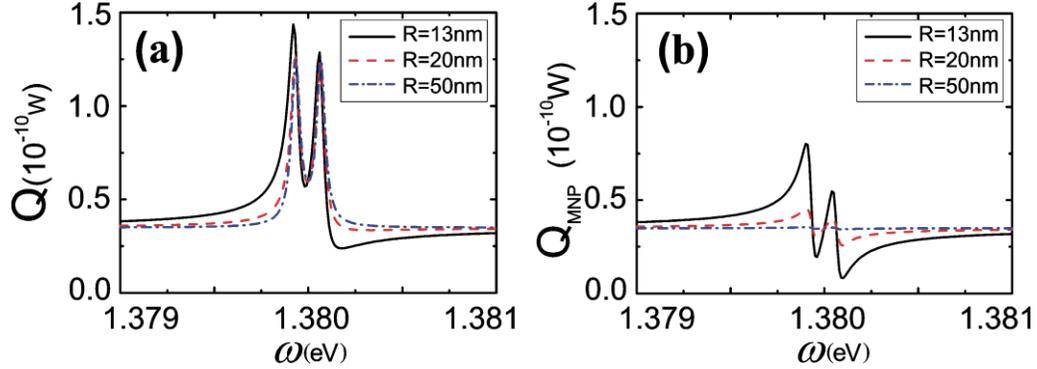

Fig. 3 (Color online). Energy absorption spectrum versus the frequency of the exciting laser field, $\omega$, for different interparticle distances, R. (a) Totle energy absorption rate of the MNP-SQD hybrid system. (b) Energy absorption rate in MNP. Here, we set I=1W/cm$^2$, $r$=0.65nm, $a$=3nm, $S_\alpha^{12}$=2 and $S_\alpha^{13}$=2.

We can also consider the energy absorption spectrum versus the frequency of the exciting laser field, $\omega$, for different interparticle distances, R. As shown in Figs. 3(a) and 3(b), the energy absorption spectrum appears a symmetric peak for large interparticle distances. As the interparticle distance is decreased, there appears asymmetric Fano lineshape due to the interaction between an exciton and plasmon. In our numerical calculations, we set I=1W/cm$^2$, $r$=0.65nm, $a$=3nm, $S_\alpha^{12}$=2 and $S_\alpha^{13}$=2. According to our consideration, asymmetric Fano lineshape becomes disappeared when the interparticle distance is larger than 50nm. Anyway, it seems to be true that with decreasing the distance between the particles, the absorption spectrum shows very different behavior for small distance.

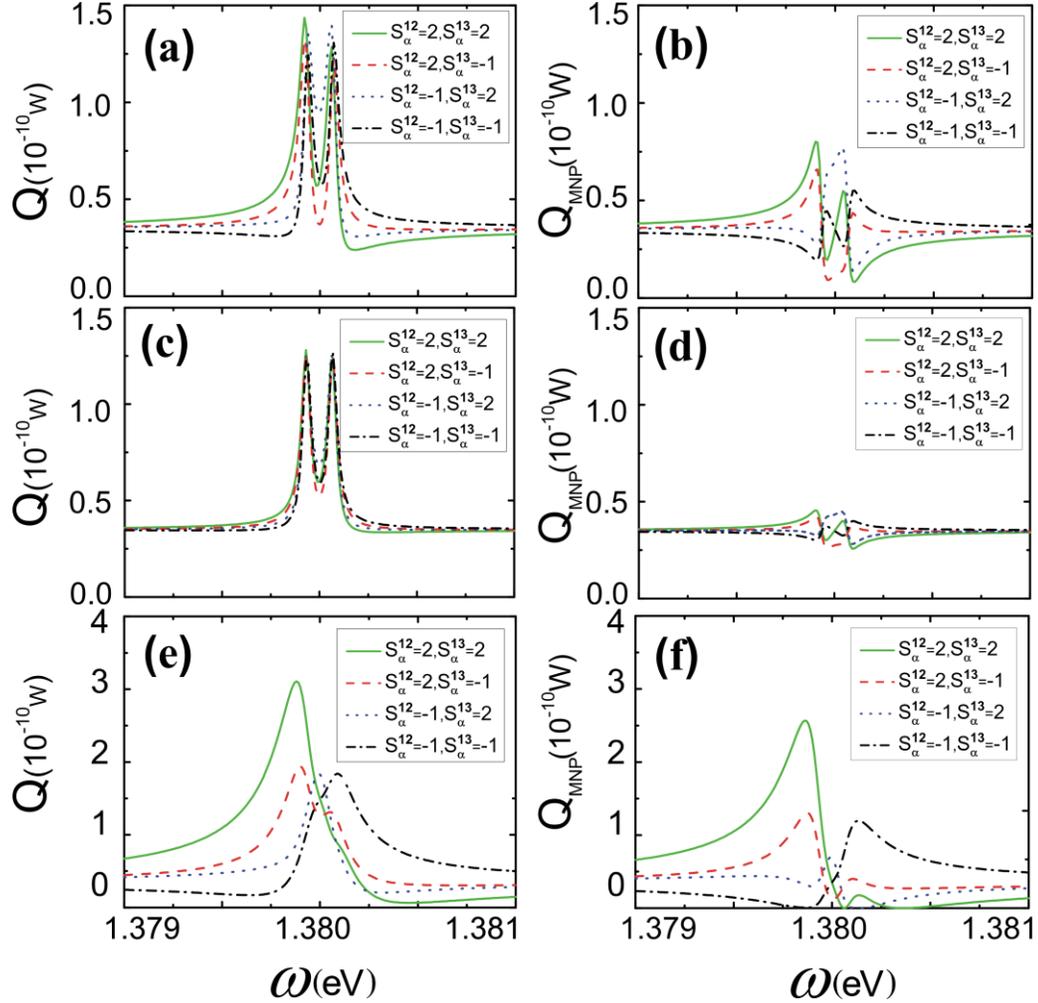

Fig. 4 (Color online). Energy absorption spectrum versus the frequency of the exciting laser field, $\omega$, for different polarization parameters. (a), (b) r=0.65nm, R=13nm. (c), (d) r=0.65nm, R=20nm. (e), (f) r=3nm, R=13nm. The rest of parameters are set as I=1W/cm$^2$ and $a$=3nm.

Figure 4 shows the energy absorption spectrum versus the frequency of the exciting laser field, $\omega$, for different polarization parameters. In Fig. 4(a) and 4(b), one can see that the energy absorption could be changed greatly for differenct polarization parameters. Especially, the energy absorption spectrum by MNP is very sensitive to the polarization parameters and exhits asymmetric Fano lineshape clearly. Figure 4(a) shows that the total absorption spectrum is almost symmetric at the interparticle distance 20nm, especially fig. 4(b) shows the absorption by metal nanoparticle is depressed strongly. We can also see the dependence of absorption spectrum on the particle size of SQD, as shown in Figs. 4(a, b) and 4(e, f). Both in the cases, the interparticle distances are set 13nm, respectively, but the sizes of SQDs are different with each other. As we can see easily from figure 4, with increasing the size of SQD, the asymmetric lineshape of the absorption spectrum becomes more prominently and the width of the absorption is enhanced strongly. What is more interesting relies on the fact that the absorption spectrum is quite sensitive to the polarization parameters, as shown in Figs. 4(e) and 4(f). We found that the absorption is enhanced very strongly when the direction of the applied electric field is parallel to the axis of the hybrid system, in terms of the polarization parameters, $S_\alpha^{12}=2$ and $S_\alpha^{13}=2$.

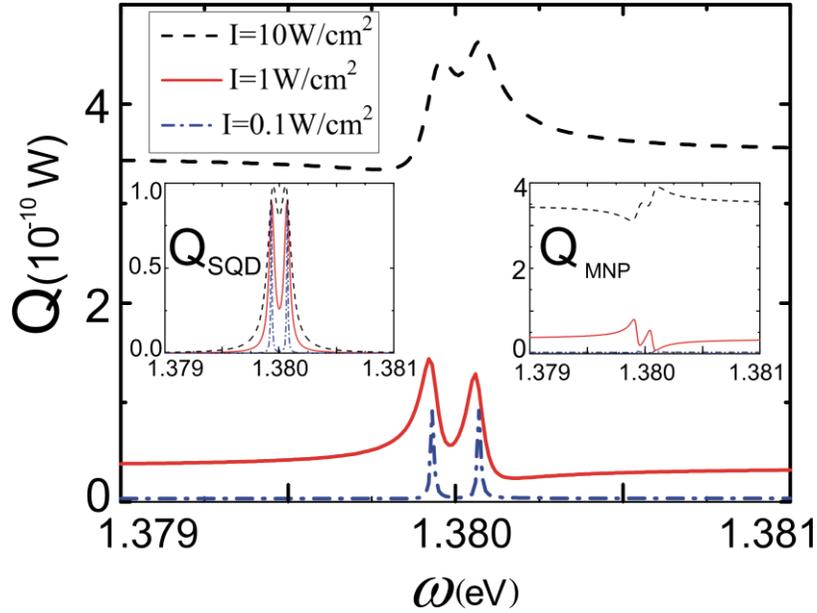

Fig. 5 (Color online). Energy absorption spectrum of MNP-SQD hybrid system versus the frequency of the exciting laser field, $\omega$, for different intensities of the laser field. The left inset is the energy absorption rate in SQD and the right inset is the energy absorption rate in MNP. We set r=0.65nm, R=13nm and $a$=3nm.

Finally, we can also consider the effect of the exciting laser field on the energy absorption spectrum of MNP-SQD hybrid system. The energy absorption spectrum of MNP-SQD hybrid system versus the frequency of the exciting laser field, $\omega$, for different intensities of the laser field was shown in Fig. 5. The left inset is the energy absorption rate in SQD and the right inset is the energy absorption rate in MNP. The result shows that with increasing the intensity of the driving field, the absorption by the hysbrid system is increased. We also found that the response of SQD on the intensity of the applied field becomes weak as the intensity of the applied field increase as shown in the left inset in Fig. 5. When I=10W/cm², the absorption spectrum of SQD doesn't exhibit two resonant peaks, but a single peak. However, the absorption spectrum by MNP is very distinct shown in the right inset in Fig. 5. The result also exhibts the Fano lineshape of the absorption spectrum by the hybrid system.

**4. Conclusion**

In conclusion, we have studied theoretically the optical properties of MNP-SQD hybrid nanosystem. We investigated the population and the absorption spectrum of the hybrid nanostructure and showed that the nonlinear optical response of the hybrid nanosystem can be greatly enhance or depressed due to the exciton-plasmon couplings. The results obtained here may have the potential applications of quantum information processing.

**Acknowledgments.** This work was supported by Key Project for Frontier Research on Quantum Information and Quantum Optics of Ministry of Education of D. P. R of Korea.